\documentstyle[multicol,aps,amsmath,epsfig]{revtex}

\begin{document}

\draft
\title{Crossover from Fragile to Strong Glassy Behaviour in
Kinetically Constrained Systems}

\author{Arnaud Buhot\cite{emailAB}
and Juan P. Garrahan\cite{emailJPG}}

\address{Theoretical Physics, University of Oxford, 1 Keble
Road, Oxford, OX1 3NP, U.K.}

\date{April 18, 2001}

\maketitle

\begin{abstract}
We show the existence of fragile-to-strong transitions in kinetically
constrained systems by studying the equilibrium and out-of-equilibrium
dynamics of a generic constrained Ising spin chain which interpolates
between the symmetric and fully asymmetric cases.  We find that for
large but finite asymmetry the model displays a crossover from fragile
to strong glassy behaviour at finite temperature, which is controlled
by the asymmetry parameter. The relaxation in the fragile region
presents stretched exponential behaviour, with a temperature dependent
stretching exponent which is predicted. Our results are confirmed by
numerical simulations.
\end{abstract}


\vspace{-0.3cm}

\begin{multicols}{2}
\narrowtext

Glasses are everywhere in nature. It is difficult to find a liquid
which when supercooled does not form the amorphous, microscopically
disordered solid we call a glass. The salient feature of supercooled
liquids is their dramatical slowing down with decreasing temperature,
signaled by the increase of their relaxation times and viscosities by
several orders of magnitudes in a temperature range of a few
decades. For reviews see \cite{Angell,Angellplus,Debe}. 

Given the generic nature of the glassy state, universal principles for
the classification of glass-forming materials are of paramount
importance. Central to this is the concept of fragility
\cite{fragile,Angell}, which measures the speed with which viscosity
and relaxation times grow as a system approaches the glass
transition temperature.  Liquids which display Arrhenius behaviour,
that is, the logarithm of their viscosity or relaxation time grows
linearly with inverse temperature, are classified as `strong'
\cite{fragile,Angell}, as for example network liquids like SiO$_2$ and
GeO$_2$, and define one of the extremes of the classification by
fragility.  Most liquids, however, behave in a non-Arrhenius manner,
and the larger the departure from Arrhenius behaviour, the more
`fragile' \cite{fragile,Angell}, the most fragile liquids, which
define the second extreme in fragility, being polymeric in nature. An
exception to this classification is supercooled water \cite{Ito}: at
temperatures close to its melting point it behaves as extremely
fragile, while near the glass transition it is very strong.
Supercooled water has a fragile-to-strong transition.

Some of the simplest systems which display the slow cooperative
relaxation characteristic of glasses are the facilitated kinetic Ising
models, first introduced by Fredrickson and Andersen
\cite{Fredrickson}, in which glassiness is not a consequence of either
disorder or frustration in the interactions, but of the presence of
kinetic constraints in the dynamics of the system.  Depending on
whether the constraints are isotropic \cite{Fredrickson}, or fully
directed \cite{Jackle}, these models may behave as strong or fragile
glasses, and the low temperature dynamics can be understood in terms
of activation over energy barriers
\cite{Schulz,Mauch,Sollich,Crisanti}. They are particularly useful in
the study of activated processes, which become highly relevant for
supercooled liquids near the glass transition, but are not taken into
account by approximations like mode-coupling theory \cite{Gotze} or
mean-field models \cite{Kirkpatrick,Bouchaud}.

The purpose of this Letter is to show that fragile-to-strong
transitions also occur in kinetically constrained systems.  We prove
this for the simplest case of the one-spin facilitated Ising chain. We
study a generalization of the kinetically constrained Ising chain
which interpolates between the cases of symmetric \cite{Fredrickson}
and fully asymmetric constraints \cite{Jackle}. We show that for large
but finite asymmetry the model displays a fragile-to-strong crossover
at finite temperature. The crossover temperature is controlled by the
asymmetry parameter, which also determines the largest timescale and
energy barrier of the problem. The relaxation in the fragile region
presents stretched exponential behaviour, with a temperature dependent
stretching exponent which is obtained analytically. We performed
extensive numerical simulations to confirm our results.

We consider a chain of Ising spins $\sigma_i \in \{0,1\}$ $(i=1, \dots,
N)$, with periodic boundary conditions, and Hamiltonian $H = \sum_i
\sigma_i$. The dynamics is restricted to single flips of spins which
have at least one nearest neighbour in the up state.  The rates for
the possible transitions are in general different depending on whether
the up neighbour is on the right or left, and are given by:
\begin{equation}
11 \stackrel{b}{\longrightarrow} 01, \,\,\,
01 \stackrel{b \epsilon}{\longrightarrow} 11, \,\,\,
11 \stackrel{1-b}{\longrightarrow} 10, \,\,\,
10 \stackrel{(1-b) \epsilon}{\longrightarrow} 11,
\label{rates}
\end{equation}
where $b \in [0,1]$ and $\epsilon \equiv \exp(-1/T)$. Detailed balance
is obeyed, and the stationary distribution is the Boltzmann
distribution at temperature $T$ for the Hamiltonian $H$. The parameter
$b$ sets the degree of asymmetry of the kinetic constraints. The
limiting values $b=1/2$ and $b=0$ (or $1$) correspond to the
Fredrickson--Andersen (FA) model \cite{Fredrickson} and the
asymmetrically constrained Ising chain (ACIC) \cite{Jackle},
respectively.  

Due to the non-interacting nature of the Hamiltonian the
thermodynamical properties of the model are trivial, and are the same
for any $b$. The energy density is given by the concentration $c$ of
up spins (or `defects'), which in equilibrium becomes $c_{\rm
eq}=\epsilon/(1+\epsilon)$. At low temperatures $c$ is very small, and
since defects facilitate the dynamics, the system slows down: isolated
up spins are locally stable and the system has to overcome energy
barriers to evolve.

In contrast with the statics, the low temperature dynamics depends
strongly on the value of $b$ (except at $T$ strictly zero
\cite{Crisanti}). In the symmetric limit $b=1/2$, which corresponds to
the FA model, isolated defects can diffuse to the left (resp.\ right)
by means of processes (ii) and (iii) [resp.\ (iv) and (i)] of Eq.\
(\ref{rates}). Each move requires the temporary creation of one
defect, so there is a single activation barrier to diffusion $\Delta
E=1$. This constancy of the energy barriers implies that relaxation
times follow the Arrhenius law $\tau_{\rm FA} \sim \exp(\Delta E/T)$,
characteristic of strong glass behaviour \cite{Schulz,Crisanti}. The
decay of the concentration of up spins, in the out of equilibrium
regime, is well approximated by the coagulation process $A+A
\rightarrow A$, which means that the typical lengthscale grows as $l
\sim (t/\tau)^{1/2}$ \cite{Schulz}.  The situation in the asymmetric
limit $b=0$ or $1$, which corresponds to the ACIC model \cite{Jackle},
is very different. Here the activated diffusion mechanism of the
symmetric case is absent: a defect at a distance $2^{n-1} < d \leq
2^n$ from the nearest defect (in the direction of the constraint) has
to cross a barrier $\Delta E = n$ to move \cite{Sollich}, i.e.,
barriers grow with the logarithm of the size of relaxing regions. This
means that typical lengthscales grow as $l \sim t^{T \ln 2}$, which
leads to a relaxation time at low temperatures of the form $\tau_{\rm
ACIC} \sim \exp(1/T^2 \ln 2)$ \cite{Sollich}.  This is the B\"assler
law \cite{Bassler,Angellplus} used as an alternative to the
Vogel-Fulcher equation \cite{Angell,Angellplus} to represent fragile
behaviour. The absence of a finite temperature singularity is
consistent with the trivial statics of the ACIC model.

We now consider the behaviour at intermediate values of the asymmetry
$b$. We first focus on the relaxation towards equilibrium after a
quench from infinite temperature. When $b$ is not far from the
symmetric limit $b=1/2$, the rates of Eq.\ (\ref{rates}) for reactions
to the left and right are comparable and the symmetric diffusive
mechanism is still effective: the behaviour is essentially that of the
FA model. The region of large but finite asymmetry, that is, when $b$
(or $1-b$) is small, is more interesting. In this case rates (i) and
(ii) of Eq.\ (\ref{rates}) are very much suppressed respect to rates
(iii) and (iv) (or vice-versa), the system can only make use of the
asymmetric mechanism to relax, and behaves like the ACIC. The
timescales associated with the asymmetric process, however, grow as
the relaxing regions become larger: the timescale for relaxation of a
region of length $2^{n-1} < d \leq 2^n$ is $\tau_A^{(n)} \sim
\epsilon^{-n}$, and the dynamics takes place in stages labeled by $n$
\cite{Sollich}. The timescale for the symmetric process, $\tau_S \sim
[b(1-b) \epsilon]^{-1}$, is large, but not infinite, except in the
ACIC limit, and in contrast to the asymmetric one does not grow with
increasing length. This means that eventually $\tau_S$ becomes
comparable to $\tau_A^{(n)}$ for some value of $n$. This defines the
last stage of asymmetric relaxation $n^*$ before the system switches
to symmetric behaviour:
\begin{equation}
n^* \sim 1 - T \ln [b(1-b)] .
\label{nstar}
\end{equation}

\begin{figure}[t]
\begin{center}
\epsfig{file=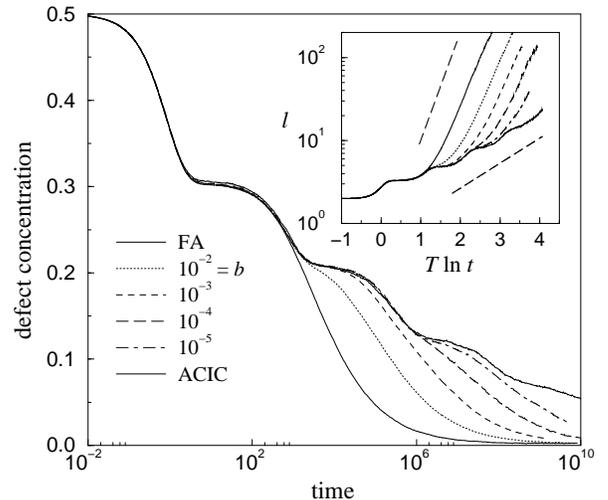, width=3.1in}
\caption
{ Defect concentration $c$ as a function of time $t$ after a quench
from infinite temperature to $T=1/6$, for values of the asymmetry
$b=1/2$ (FA model), $10^{-2}$, $10^{-3}$, $10^{-4}$, $10^{-5}$, and
$0$ (ACIC model). Inset: typical lengthscale $l \equiv 1/c$ against
scaled time variable $T \ln t$. The upper dashed line corresponds to
$t^{1/2}$ and the lower one to $t^{T \ln 2}$.  Simulations were
performed using a continuous time Monte Carlo/Metropolis algorithm
$[16,17]$ for a system of $10^5$ spins.  }
\label{energy}
\end{center}
\end{figure}

In Fig.\ \ref{energy} we show the decay of defect concentration $c$
(the energy density of the system) as a function of time $t$ after a
quench from an initial state at $T_0=\infty$ to a low temperature
$T=1/6$, for several values of the asymmetry $b$. For the $b=0$ ACIC
limit, at least four plateaus in the concentration are visible, which
correspond to the different stages in the dynamics.  The number of
plateaus decreases with increasing $b$ in accordance with Eq.\
(\ref{nstar}): there are three plateaus for $b=10^{-5}$ ($n^*=2.9$),
two for $b=10^{-4}$ and $10^{-3}$ ($n^*=2.5$ and $2.2$), and only one
for $b=10^{-2}$ and $1/2$ ($n^*=1.7$ and $1.1$), the latter being the
FA limit. The Inset shows how the typical lengthscale $l \equiv 1/c$
grows with time. While in the ACIC case it follows $t^{T \ln 2}$
(lower dashed line), for all the nonzero values of $b$ the system
eventually switches to the diffusive $t^{1/2}$ behaviour of the FA
case (upper dashed line).

Let us turn to the implications that a finite asymmetry has on the
relaxation time of the system. For arbitrary $b$, the symmetric and
asymmetric processes compete. The relaxation timescale is given by
$\tau \sim \left( \tau_S^{-1} + \tau_{\rm ACIC}^{-1} \right)^{-1}$,
assuming that the corresponding rates add up.  The factor $b(1-b)$ in
the symmetric relaxation timescale $\tau_S$ can be interpreted as an
entropic barrier: $\tau_S \sim \exp\{[\Delta E - T \ln b(1-b)]/T\}$,
where $\Delta E=1$.  Notice that this `free energy' barrier is
precisely $n^*$ of Eq.\ (\ref{nstar}).  Thus, the suppression of the
symmetric mechanism due to $b$ decreases with decreasing
temperature. The consequence of this on the relaxation time for a
fixed value of $b$ is the following.  At higher temperatures the
asymmetric process dominates, $\tau \sim \tau_{\rm ACIC}$, and the
system displays fragile relaxation. At lower temperatures, the
symmetric process becomes dominant, $\tau \sim \tau_S$, and the
behaviour is strong. The crossover temperature for this
fragile-to-strong transition is determined by $b$,
\begin{equation}
T_c \sim \frac{1 - \sqrt{1 - 4 \ln [b(1-b)] / \ln 2}}
{2\,\ln [b(1-b)]},
\label{tc}
\end{equation}
corresponding to $\tau_S \sim \tau_{\rm ACIC}$.

\begin{figure}[t]
\begin{center}
\epsfig{file=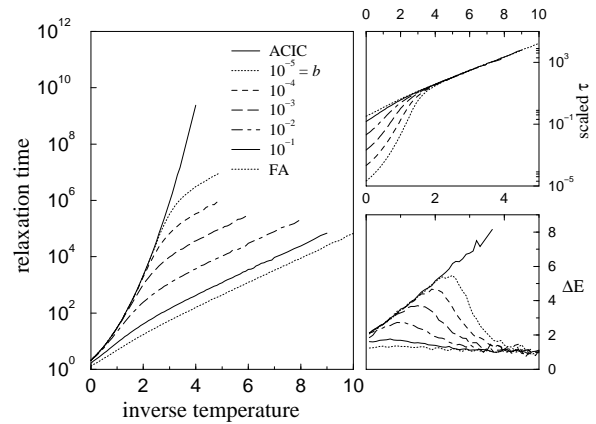, width=3.1in}
\caption
{ Equilibrium relaxation time $\tau$ as a function of inverse
temperature $1/T$, for values of the asymmetry $b=1/2$ (FA model),
$10^{-1}$, $10^{-2}$, $10^{-3}$, $10^{-4}$, $10^{-5}$, and $0$ (ACIC
model). Top-right: rescaled time $b(1-b)\tau$ against $1/T$.
Bottom-right: effective activation barrier $\Delta E$ as a function of
$1/T$. Simulations were performed for system sizes which
varied from $10^5$ for high temperatures to $5 \times 10^6$ for low
ones, and data points were averaged over twenty runs.}
\label{times}
\end{center}
\end{figure}

In Fig.\ \ref{times} we show the equilibrium relaxation time $\tau$ as
a function of inverse temperature $1/T$.  Given that the static
properties of the system are known exactly it is simple to construct
low temperature equilibrium configurations.  This allows to study the
equilibrium dynamics down to really low temperatures in contrast with
most glassy systems where only out of equilibrium quantities are
accessible. We obtain the relaxation time through the connected
equilibrium autocorrelation function, $C(t) \equiv N^{-1} \sum_i
\langle \sigma_i(t) \sigma_i(0) \rangle - c_{\rm eq}^2$, where $\tau$
is defined by $C(\tau) = e^{-1} C(0)$. Alternative definitions of the
equilibrium relaxation timescale give similar results. The figure
shows the following features: (i) in the $b=0$ ACIC limit, the
relaxation time has the fragile behaviour $\tau_{\rm ACIC}$ for all
temperatures, as expected; (ii) for small $b$, $\tau$ crosses over
from fragile $\tau_{\rm ACIC}$ at higher temperatures to strong
$\tau_S$ behaviour at lower ones, the crossover taking place around
$T_c$ given by Eq.\ (\ref{tc}) (e.g., $1/T_c = 3.2$ for $b=10^{-5}$
and $1/T_c = 2.9$ for $b=10^{-4}$); (iii) for larger $b$ the fragile
region shrinks, and disappears completely at the FA limit $b=1/2$. The
displacement of the curves in the strong regime is due to the entropic
barrier. Rescaling the relaxation time by $\tau \rightarrow b(1-b)
\tau$ makes them collapse, as shown in the top-right panel of Fig.\
\ref{times}. The bottom-right panel gives the effective activation
barrier $\Delta E \equiv d \ln \tau / d(1/T)$. The crossover here
appears as a change from the linear growth in the ACIC to the constant
barrier of the FA model. The crossover is sharper the smaller $b$.
The high temperature behaviour of $\tau$ is exponential in $1/T$,
which gives the offset in the straight line for $b=0$, and becomes
irrelevant at low temperatures.  The slope of $\Delta E$ is about
$1.7$ rather than $2/\ln 2$. This difference is due to the
non-exponential nature of the autocorrelation, which does not affect
the $1/T^2$ behaviour of the log of $\tau$.

The fragile-to-strong crossover can also be observed in the behaviour
of equilibrium dynamical quantities. The simplest one to study is the
equilibrium persistence, $P(t) \equiv N^{-1} c^{-1} \sum_i \langle
\prod_{t'=0}^t \sigma_i(t') \rangle$, which measures the fraction of
defects of the initial configuration which have never flipped between
times $0$ and $t$. It is closely related to the equilibrium
autocorrelation $C(t)$, but is free from the problem of the recurrence
of defects which makes the analysis of the latter more tricky,
particularly in one dimension.

As mentioned before, in the asymmetric limit, the
probability to flip a defect depends on the distance to the nearest
defect in the direction of the constraint. In equilibrium, the
probability distribution of these distances is independent of time.
The persistence may be then approximated by the sum of the independent
exponential relaxation of defects at different distances from their
neighbours, 
$P(t) \sim c_{\rm eq} \, e^{-t/\tau_0} + 
\sum_{n=1}^{\infty} p_n e^{-t/\tau_A^{(n)}}$,
where $\tau_0 \sim 1$ is the timescale associated with the initial $T$
independent transient, and $p_n = (1-c_{\rm eq})^{2^{n-1}}-(1-c_{\rm
eq})^{2^n}$ is the probability for a defect to have a chain of spins
$0$ of length $2^{n-1} \leq d < 2^n$ next to it, which takes into
account the fact that the relaxation time for the corresponding
distances is $\tau_A^{(n)}$. The equilibrium condition is crucial to
assume an independent relaxation of the different lengthscales. If the
initial configuration is an out of equilibrium one, the probability
distribution of distances evolves in time and the approximation
considered is no longer valid. For short times the persistence is
dominated by the fastest exponential decay, leading to $- \ln P \sim
t/\tilde{\tau}$ for low temperatures, where $\tilde{\tau} \equiv
\tau_0/c_{\rm eq}$ defines the timescale for short times. The long
time behaviour may be estimated replacing the sum by an integral which
is evaluated in the saddle point approximation, in a manner similar to
that of Ref.\ \cite{Palmer}.  As a result we obtain a stretched
exponential $P(t) \sim \exp \left[ - \left(t/\tau_{\rm ACIC}
\right)^{\beta} \right]$, with a stretching exponent
\begin{equation}
\beta = (1 + 1/T \ln 2)^{-1}.
\label{beta}
\end{equation}
Notice that this is an alternative method to obtain $\tau_{\rm ACIC}$
to the one of Ref.\ \cite{Sollich}.  In the symmetric limit the
persistence is simply the sum of the relaxation of defects with and
without an up neighbour. At low temperatures it reads
$P(t) \sim 2 \, c_{\rm eq} \, e^{-t/\tau_0}  
+ (1-2 \, c_{\rm eq}) \  e^{-t/\tau_S}$. 
The small time behaviour is similar to the asymmetric case, while the
long time one is given by $P(t) \sim \exp(-t/\tau_S)$.

\begin{figure}[t]
\begin{center}
\epsfig{file=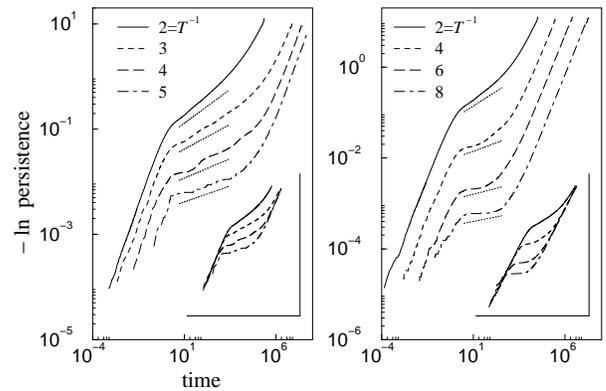, width=3.1in}
\caption
{ Persistence $P$ as a function of time $t$, for $b=10^{-5}$ (left
panel) and $b=10^{-3}$ (right panel). Notice that we plot $- \ln P(t)$
in a log-log scale. The dotted lines are the expected stretching
exponents $\beta$. Insets: same as main panels with time rescaled as
$t \rightarrow t \epsilon$. Details of simulations are the same as in
Fig. \ref{times}.}
\label{pers}
\end{center}
\end{figure}

The behaviour of the persistence for $b=10^{-5}$ and $b=10^{-3}$ at
different temperatures is shown in Fig.\ \ref{pers}.  We present the
data in a double log scale for $P$ and a log scale for $t$ to display
the different stretching exponents. As in the case of the out of
equilibrium relaxation, the decay is first dominated by the asymmetric
process corresponding to the fragile regime. For short times it is
exponential with a characteristic timescale $\tilde{\tau} \sim
\epsilon^{-1}$, as described above.  At longer times we see a change
of slope in the plot, which corresponds to stretched exponential
behaviour, with an exponent given by Eq.\ (\ref{beta}). The stretching
region is larger for smaller $b$, as expected. The fragile-to-strong
crossover then takes place, and the persistence becomes exponential
again, now with a timescale $\tau_S$. In the Insets we rescale time by
a factor of $\epsilon$ to superimpose the curves in the exponential
regimes of short and long times. The two limiting lines correspond to
$\exp(-t/\tilde{\tau})$ and $\exp(-t/\tau_S)$.

We conclude with a comment on the explicit spatial asymmetry in the
definition of the model studied here.  It seems that to obtain other
than strong behaviour in systems with kinetic constraints it is
necessary to consider cases in which spatial isotropy is explicitly
broken, like in the ACIC \cite{Jackle} or its generalizations, which
is a rather unphysical feature. This is also the case of systems with
interactions, but which display a dynamical behaviour similar to the
spin facilitated models \cite{tritri}. The system considered in this
work, however, can be defined in an alternative but explicitly
spatially symmetric formulation, which also clarifies the relation
between the asymmetry $b$ and the timescale and lengthscale at which
the fragile-to-strong crossover takes place \cite{Chandler}. Consider
$b$ as a collective field $b(t)$ which takes values $0$ and $1$ with a
characteristic timescale for flipping $\tau_b$. The simplest physical
choice for this dynamics would be a Poisson process. $b(t)$ may also
have a spatial dependence, but on lengthscales larger than the typical
ones for the spin system. Since $\langle b(t) \rangle = 1/2$ spatial
symmetry is unbroken. The instantaneous value of $b(t)$ is either $0$
or $1$, so for times smaller than $\tau_b$ the behaviour is that of
the asymmetric case. For times much larger than $\tau_b$ the system
effectively sees the average of $b(t)$ and the behaviour is the
symmetric one. $\tau_b$ sets the timescale for the fragile-to-strong
crossover. This argument is easily generalized to higher dimensions by
considering a collective vector field instead.

\vspace{0.2cm}

The authors would like to thank David Chandler, Andrea Crisanti,
F\'elix Ritort, Andrea Rocco, and Peter Sollich for useful
discussions. The work of AB is supported by EU Grant No.\
HPMF-CT-1999-00328 and that of JPG by a Violette and Samuel Glasstone
Research Fellowship.

\end{multicols}
\end{document}